\documentstyle[preprint,aps]{revtex}
\tightenlines 
\begin{document}
 \title{  Magnetoconductance   of
 parabolically confined\\  quasi-one dimensional channels}  

\author{  S. Guillon \cite{sguill}, P. Vasilopoulos\cite{takis},
and C. M. Van Vliet\cite{vanvliet}\\
\ \\}

\address{ Concordia University, Department of Physics,\\
1455 de Maisonneuve Ouest, Montr\'{e}al, Qu\'{e}bec, Canada, H3G 1M8\\
\ \\
\cite{vanvliet} Department of Physics, University of Miami,\\
PO Box 248046, Coral Gables, FL 33174, USA}
\date{\today} \address{} \address{\mbox{}} \address{\parbox{14cm}{\rm 
\mbox{}\mbox{}\mbox{} Electrical conduction is studied along 
parabolically confined quasi-one dimensional channels, in the 
framework of a revised linear-response theory, for elastic scattering.  For
zero magnetic field an explicit multichannel expression for the 
conductance is obtained that agrees with those of the literature.  A 
similar but new multichannel expression is obtained in the presence of 
a magnetic field $B||\hat{z}$ perpendicular to the channel along the x
axis. An explicit connection is made between the characteristic  time for the
tunnel-scattering process and the transmission and reflection coefficients that
appear in either expression. As expected, for uncoupled channels  the finite field
expression gives the complete (Landauer-type)  conductance of N parallel 
channels, a  result that has  not yet been reported in the literature.
In addition, it accounts explicitly for the Hall field and the confining
potential and is valid, with slight modifications, for tilted magnetic fields in
the (x,z) plane.}}

\address{\mbox{}}
\address{\parbox{14cm}{ \rm PACS 73.40.Gk, 73.23.-b, 73.23.Ad}}
\maketitle
\clearpage
\noindent

\section{ Introduction}

The observation of the conductance quantization \cite{1} more than a 
decade brought new attention to Landauer's formula \cite{2} for the 
conductance of single-channel one-dimensional electronic systems and to 
its multichannel version derived in Ref. \cite{3} from arguments similar to those used 
by Landauer.  The single-channel formula \cite{4} and a modified version of
it \cite{4a}  have been 
derived from linear-response theory.  Slight variations 
between different results were a source of discussion \cite{4b} and crucial 
importance was given to the conditions of measurement.  It was established 
that four-probe measurements do not give the same answer as the two-probe
ones \cite{5}.  For a review of the subject we refer the
reader to Refs.  \cite{5} and \cite{6}.

The conductance has also been studied in the presence of a magnetic 
field.  The two-probe formula and its generalization have been found 
to hold.  It was derived again using linear-response theory  
\cite{7}.  The Onsager's relation, relating the symmetry of the 
conductance upon changing the direction of the magnetic field, was 
verified.  For the four-probe measurement it was realized \cite{8} and 
confirmed theoretically \cite{9} and experimentally \cite{9a} that the conductance can be asymmetric 
under reversal of the magnetic field. 
  
 As noted by the authors of Ref.  \cite{3} their multichannel formula 
 does not reduce, for uncoupled channels, to that of Ref.  \cite{2}.  
 This drawback  results from their assumption that all channels 
 originating from the reservoirs have the same electrochemical 
 potential regardless of their velocities.  In a recent Ph. D. thesis,
 Ref.  \cite{10}, completed under the direction of one of us (CMVV), a 
 multichannel formula, free from this drawback, has been derived for 
 zero magnetic field.  
 
 In this work, following Ref.   \cite{10}, we derive a rigorous 
 multichannel conductance formula in the presence of a magnetic field 
 from a revised linear-response theory.  As in almost all works of the
 literature, it is valid for elastic scattering, i.e., in mesoscopic conductors.  The formulation 
 shows explicitly the cancellation in the product of the velocity with 
 the quasi-one-dimensional density of states in the current carried by 
 a channel or mode and therefore reflects some of the intuition of 
 the original work \cite{3}.  The formula is made very explicit for 
 parabolically confined quasi-one-dimensional channels.  This type of 
 confinement allows us to easily include the Hall field which 
 simulates the electron-electron interaction in a mean-field sense 
 \cite{11}.  We also consider the case of tilted magnetic fields.

 In Sec.  II we present a general formula for the conductivity and 
 give the related one-electron characteristics.  In Sec.  III we 
 evaluate the conductance using a scattering formulation and present 
 various limits. Finally in Sec.  IV we present a discussion of the results.
 
\section{Expression for the conductivity }

\subsection{New linear-response expressions}

In order to explain our approach  we first present some general 
results, in line with those from Refs. \cite{10} and \cite{11a}, which  will be
used to derive a general expression for the magnetoconductance.  The model of
the conductor or sample we use is illustrated in Fig.  1.  It consists of two
perfect leads (reservoirs) with random scattering centers in the middle.  The
longitudinal electric field representing the potential difference is applied in
the inhomogenous part.  A magnetic field {\bf B} is applied along the z
axis ($ \vec{B} = -B \hat{z}$).

The many-body Hamiltonian that enters von Neuman's equation is
\begin{equation}
 H_{tot}(t) = H_{0} + W(t) + H^{I},
 \label{eq:nini}
\end{equation}
where $H^{I}$ represents the scattering or perturbation and 
$W(t)$ the external force.
The free-electron part $H_{0}$ will be specified later for the geometry 
of Fig. 1. For elastic scattering the equation for the {\it many-body}  density
 operator  can be 
transformed to a similar  one for the {\it one-body density} operator $ \rho(t)$.  
The latter is the sum of 
the unperturbed, Fermi-Dirac operator $f(h)$ and of 
the perturbation operator $\tilde{\rho}(t)$, i.e., $ \rho(t) = f(h)+ 
\tilde{\rho}(t)$.  For linear responses and with the initial condition 
$ \tilde{\rho}(0) =0$ the equation for $\tilde{\rho}(t)$ reads

 \begin{equation}
(\partial\tilde{\rho} (t)/\partial t)  + i  \tilde{{\cal L}} 
\tilde{\rho} (t)
 = -(i/\hbar) [\tilde{w}(t),f(h)],
 \label{eq:yxzw}
 \end{equation}
 where $ \tilde{{\cal L}}\bullet \equiv (1/\hbar)[ h(t), \bullet ]$ and
 $\bullet$ stands for an arbitrary one-body operator. 
 The solution is found using the resolvent of $ \tilde{{\cal L}}$, i.e., the
  Laplace transform of Eq. (\ref{eq:yxzw}). 
 In the Laplace domain Eq. (\ref{eq:yxzw}) reads
 %
  \begin{equation}
    \tilde{\rho}(s) = -\frac{i}{\hbar} 
                    \frac{1}{s+ i \tilde{\cal{L}}}
                    [\tilde{w}(s),f(h)]
  \end{equation}
 In a representation in which $H_{0}$
 is diagonal so is  its one-body counterpart $h_0$. In this representation
 the operator $\tilde{\rho}$ has a diagonal
 ($\tilde{\rho}_{d}$) and a nondiagonal ($\tilde{\rho}_{nd}$) part, $\tilde{\rho} = 
\tilde{\rho}_{d} + \tilde{\rho}_{nd}$. Substituting this in Eq. (3) 
and acting on it with diagonal ($\cal{P}$) and nondiagonal ($1-\cal{P}$)
projection superoperators leads
to two coupled equations, one for $\tilde{\rho}_{d} $ and one for $\tilde{\rho}_{nd}$. 
 %
 The steady state solution of these equations is represented by the limit $ t \rightarrow 
 \infty$.  In  Laplace domain this is equivalent to the limit $s \rightarrow 0+$.  
 
 The result obtained for the diagonal part $\tilde{\rho}_{d}$ of the density 
 operator, the only one pertinent to the conductance, is

\begin{equation}
   \tilde{\rho}_d= -\frac{i}{\hbar} \tilde{\Lambda}^{-1} \Gamma 
\sum_{\alpha\beta}[w,f(h)]_{\alpha\beta}
	\vert \psi_\alpha\rangle \langle \psi_{\beta} \vert.
\end{equation}
Here $\tilde{\Lambda}$ and $\Gamma$ are superoperators  
associated with the transitions caused by the perturbation $h^I$. They are
given by 
$\tilde{\Lambda}={\cal P}{\cal L}^{1} [1/(i {\cal
L}+0^+)]{\cal L}^1$ and 
	$\Gamma={\cal P}\left [ 1-{\cal L}^{1}[1/(i{\cal L}+0^{+})]{\cal L}^{1}
	\right ]$ with
$\cal L$ and ${\cal L}^{1}$  defined by 
${\cal L} \bullet \equiv  [H, \bullet ]/\hbar$
 and $ {\cal L}^1 \bullet \equiv
    [V, \bullet ]/ \hbar$.  $\tilde{\Lambda}$ is the one-particle scattering operator.
The one-body analog $w$ of $W$ is related to the  
 electric field by $e{\bf E({\bf r})}= - \nabla w ({\bf r})$. 
Further, $|\psi_{i}\rangle$ are the eigenstates of $h=h_0+h^I$, i.e., 
$h|\psi_{i}\rangle = E_{i} |\psi_{i}\rangle $.   The operator $\Gamma$ doesn't 
affect the sum and the number $[w,f(h)]_{\alpha\beta}$.
Using the relation 
$  \Gamma\vert \psi_\alpha\rangle \langle \psi_{\beta} \vert = 
\vert \varphi_\alpha\rangle \langle \varphi_{\beta} \vert 
\delta_{\alpha\beta}$, where $|\varphi_\alpha\rangle$ is the eigenstate of 
$h_0$ and  
 
\begin{equation}
\langle\psi_{\alpha}\vert[w,f(h)]\vert \psi_{\beta}\rangle
= -i\hbar \ \frac{ f(\epsilon_{\beta})-f(\epsilon_{\alpha}) 
}{\epsilon_{\beta}-\epsilon_{\alpha}}  
\int_{V_0}d{\bf r}'E( {\bf r}')\langle\psi_{\alpha}\vert 
j({\bf r}')\vert\psi_{\beta}\rangle , 
\end{equation}
with $ f(h)\psi_{i}=f(\epsilon_{i})$, we have
\begin{equation}
\langle\varphi_{\theta}\vert\tilde{\rho_d}\vert\varphi_{\gamma}\rangle 
=-
\sum_{\alpha\beta}
     \langle\varphi_{\theta}\vert\tilde{\Lambda}^{-1}\vert 
\varphi_\alpha\rangle
     \delta_{\beta\gamma}
      f'(\epsilon_{\alpha})\delta_{\alpha\beta}
\int_{V_0}d{\bf r}'\langle\psi_{\alpha}\vert 
j({\bf r}')\vert\psi_{\beta}\rangle E({\bf r}'),
\label{densite}
\end{equation}
%
where $V_0$ is the volume. The current density is
\begin{equation}
J({\bf r}) =  Tr\{ j({\bf r}) \tilde{\rho}_d \} \nonumber \\
     =  \sum_{\gamma\theta}\langle \varphi_{\gamma} 
           \vert j({\bf r})\vert \varphi_{\theta} \rangle
           \langle \varphi_{\theta} \vert  \tilde{\rho}_d 
           \vert \varphi_{\gamma}\rangle. 
\end{equation}
Substituting Eq. (\ref{densite}) into Eq. (7) and
comparing the result with the general expression

\begin{equation}
   J({\bf r})=\int_{V_0}d{\bf r}'\sigma({\bf r},{\bf r}')E({\bf r}')
\end{equation}
we find the following expression for the conductivity 

\begin{equation}
\sigma({\bf r},{\bf r}')\equiv\stackrel{\leftrightarrow}{\sigma}_{d}
({\bf r},{\bf r}') =  -\sum_{\gamma\theta}j_{\gamma\theta}({\bf r})
     \langle\varphi_{\theta}\vert\tilde{\Lambda}^{-1}\vert 
\varphi_\gamma\rangle
     f'(\epsilon_{\gamma})\langle\psi_{\gamma}\vert 
j({\bf r}')\vert\psi_{\gamma}\rangle,
     \label{eq:baba}
\end{equation}
where the left-right arrow indicates that $\sigma({\bf r},{\bf r}')$ is a tensor.
The conductance $G$ is given by 

\begin{equation}
G=\int_{A}\int_{A'}dA {\bf .} \stackrel{\leftrightarrow}{\sigma}_{d}({\bf r},
{\bf r}'){\bf .} dA'
     \label{cond}
\end{equation}
where $A$ and $A'$ are two suitably chosen surfaces.

 
 \subsection{One-electron characteristics}
 
 {\it Eigenfunctions and eigenvalues}. We consider an  electron gas  which 
 interacts only with impurities.
 As  shown  in Fig. 1, a magnetic field  {\bf B}=$ -B \hat{z}$ is
applied along the z axis. 
When an electric
field $\vec E_{x}$ is applied the resulting Hall field 
$E_{\perp}$ is opposite to the y axis. We consider a  parabolic confining 
potential  along the y axis, $V_{y}=m\Omega^{2} y^{2}/2$ and 
use the vector potential 
${\bf A}=By\hat{x}$. Including the field
$E_{\perp} $\cite{11c} in the one-electron   hamiltonian $h_0$  
gives

     \begin{equation}
      h_{0} = \frac{1}{2m}\left ( \vec{P} - q {\bf A} \right )^{2}  
           -q E_{\perp} y + \frac{1}{2} m \Omega^{2} y^{2}.
               \label{eq:hamiltonien} 
     \end{equation}

We attempt a solution of Eq.  (\ref{eq:hamiltonien}) in the form 
           $\varphi (x,y) = \chi (y) \exp(i k_{x} x)$ and introduce the
           variable $\xi = \hbar k_{x}/qB + q E_{\perp}/m 
\omega^{2}_{c}$, where 
  $ \omega_{c}= qB/m$ is the cyclotron frequency. Using $ \omega^{2}_{T} = \omega^{2}_{c} + 
\Omega^{2}$, $h_{0}\varphi (x,y)
  =\epsilon\varphi (x,y)$ and completing the square
  we can rewrite Eq. (\ref{eq:hamiltonien}) as
        \begin{equation}
          \frac{m \omega^{2}_{T}}{2}  ( y - 
\frac{\omega^{2}_{c}}{\omega^{2}_{T}}
           \xi  )^{2} 
          -\frac{\hbar^{2}}{2m} \chi^{''} (y) 
          = E \chi (y),
          \label{eq:yo1}
         \end{equation}
where 
 $E = \epsilon - E(k)$.
With $\zeta=[y - (\omega^{2}_{c}/\omega^{2}_{T})\xi] 
(m \omega_{T}/\hbar)^{1/2}$ 
the solution of Eq. (\ref{eq:yo1}) is $\chi_{n} (\zeta) = 
e^{-\zeta^{2}/2} H_{n} (\zeta)$, where $ H_{n} (\zeta) $ are the Hermite polynomials. 
The corresponding eigenvalues $\epsilon = E + E(k)\equiv \epsilon (k_{x}, n) $ are given by

        \begin{equation}
          \epsilon (k_{x}, n) =  (n+1/2 ) \hbar \omega_{T} 
                          + (\hbar^{2} k_{x}^{2}  \Omega^{2} 
                          -  2\omega_{C} \hbar k_{x}qE_{\perp}
                          - q^{2}E_{\perp}^{2})/2 m\omega^{2}_{T},
            \label{eq:energie} 
         \end{equation}                
where $n$ is the Landau level index.  From this expression we obtain 
the velocity ${\bf v}={\bf \nabla}_{{\bf k}} \epsilon ({\bf k})/ \hbar$
along the direction of propagation. The result is 
      \begin{equation}
       	v_{x} = (\hbar k_{x} \Omega^{2} - \omega_{c} q E_{\perp})/ m\omega^{2}_{T}. 
       \end{equation}            
 
{\it Current density}.
The current density operator is expressed in terms of the 
one-particle eigenfunctions in  matrix form. From quantum field theory 
\cite{11a} 
${\bf j}=(\hbar/i)\int \Psi^* {\bf v} \Psi d^3r$ we obtain
 
        \begin{equation}
          		{\bf j}_{ \beta\alpha} = \frac{-iq \hbar}{2m} 
                                   \left [  \varphi^{*}_{\beta} ({\bf \nabla} 
                                       \varphi_{\alpha} )
                                    - ( {\bf \nabla} \varphi^{*}_{\beta} )
                                       \varphi_{\alpha} \right ] - 
                                       \frac{q}{m} {\bf A} 
		\varphi^{*}_{\beta} \varphi_{\alpha};   
                                    \label{dc1}
         \end{equation}                           
the  term in the square brackets represents the standard ponderomotive or
diffusion current and the term $\propto {\bf A}$ a deflection due to the
magnetic field. 
We rewrite Eq. (\ref{dc1}) in terms of  the gauge-invariant derivative 
\cite{7}
$ {\bf D} = {\bf \nabla} - iq{\bf A}/m $ 
in the form ($ f \stackrel{\leftrightarrow}{D} g = f {\bf \nabla} g - g
{\bf \nabla}^{*} f $)
%
         \begin{equation}
          {\bf j}_{ \beta\alpha} = \frac{-iq \hbar}{2m}  
\varphi^{*}_{\beta}  \stackrel{\leftrightarrow}{D} \varphi_{\alpha} .
                                  \label{eq:d2ct}
         \end{equation}
%
%
%
{
The current density
in the x direction depends  on the y coordinate. It vanishes along the y
direction  
due to the parabolic potential confinement.
 
Equation (\ref{eq:d2ct}) leads to some useful properties of  the  
current density expressed in terms of the relevant eigenvalues.
For  different eigenvalues one has
          \begin{equation}
          {\bf \nabla} {\bf j}_{ \beta \alpha} = \frac{iq}{\hbar} 
(\epsilon_{\alpha} - \epsilon_{\beta})
                         \varphi_{\alpha}\varphi^{*}_{\beta}.
          \end{equation}
 On the  other hand, for eigenfunctions of the same energy
    the following properties hold 
                \begin{equation}
                       \int dy \; \overline{\varphi}^{*}_{\pm \beta} 
(\stackrel{\leftrightarrow}{D} \cdot {\bf x}) \overline{\varphi}_{\pm 
\alpha} = \frac{\pm 2m i}{\hbar} \delta_{\alpha \beta}
                       , \; \epsilon_{\beta}=\epsilon_{\alpha},
                      \label{eq:pda}
                \end{equation}

                \begin{equation}
                            \int dy \; \overline{\varphi}^{*}_{\mp 
\beta} (\stackrel{\leftrightarrow}{D} \cdot {\bf x}) 
\overline{\varphi}_{\pm \alpha} = 0
                           , \; \epsilon_{\beta}=\epsilon_{\alpha} ,
                           \label{eq:prnd}
                \end{equation}
if the current flux  is normalized instead of  the eigenfunctions as shown 
in Ref. \cite{7}. The  new normalized eigenfunction is
 
       \begin{equation}
          \overline{\varphi}
_{\pm,a} =  e^{\pm i k_{x_{a}} x} \overline{\chi}_{n_{a}, \pm k_{x_{a}} 
}(y)/ \sqrt{ \theta_{a}};
       \end{equation}
the normalization ($\int \overline{\chi}^{2} dy =1$) constant  $  \theta $ 
 has the units of velocity; it is given by
    
\begin{equation}
               \theta_{\pm a} =\left [\hbar |k_{a}|   \Omega^{2} 
                 \mp q\omega_{c}E_\perp\right  ]/m\omega^{2}_{T}    
                  = v_{\pm a}           
\end{equation}                             
Notice the difference between $v_{\pm a} $, always positive, cf. Eq. (13), and the 
velocity given by Eq. (14).

 {\it Conductivity}. In terms of the eigenfunctions  of Eq. (12) the
 conductivity reads \cite{10} 
 
\begin{equation}
\stackrel{\leftrightarrow}{\sigma}_{d} ({\bf r},{\bf r}') =  -\int f'(\epsilon_{p})
     \stackrel{\leftrightarrow}{\sigma}^{\epsilon_{p}}_{d} ({\bf r},{\bf r}') d\epsilon_{p},
\end{equation}
where
     \begin{equation}
       \stackrel{\leftrightarrow}{\sigma}^{\epsilon_{p}}_{d} ({\bf r},{\bf r}') =
\sum_{s} \delta ( \epsilon_{p} - \epsilon_{s} ) \left (
       \tilde{ \Lambda}^{-1} j({\bf r})
       \right )_{ss} j({\bf r}')_{SS}.
       \label{eq:hoco}
     \end{equation}
 Here $ f'(\epsilon_{s}) $ is the derivative of the Fermi-Dirac  function, $s 
 \equiv\{n, k_{x}\} $,  $ \varphi_{s}$ are the unperturbed states, 
 and $ \psi_{S}$ the scattering states. We have also used the 
notation $\langle\varphi_{s}|X|\varphi_{s'}\rangle=X_{ss'} $ and  
$\langle\psi_{S}|X|\psi_{S'}\rangle=X_{SS'} $ for 
the matrix elements of $X$.
%
 %
 The Dirac $\delta$ function is rewritten in terms of $ k_{x} $ using the 
property $  \delta \left ( g(k_{x} ) \right ) = \sum_{i}
                                          \delta ( k_{x} -
                                           k_{x_{i}} )/|g'(k_{x_{i}})|$, where $
g' $ is the derivative of $ g(k_{x})$ and $k_{n\pm}$ are the roots of $ g(k_{x})
= 0 $ written explicitly as
 \begin{equation}
    [\hbar^{2} \Omega^{2} k_{x }^{2}
    - 2\omega_{c}\hbar qE_{\perp}    k_{x } - q^{2}E^{2}_{\perp} ]/2m\omega^{2}_{T}
    +  (n+1/2 ) \hbar \omega_{T}  - \epsilon_{p} 
                          = 0.
   \label{eq:coul}
 \end{equation}       
The roots $k_{n\pm}$ of this quadratic equation  are 
of the form $ k_{n \pm} = [-b \pm (b^{2} - 4 a c)^{1/2}]/2a $. %
They are real-and opposite to each other-if $ c$ is negative. If this condition holds
the wave functions can propagate in different channels. For complex roots, 
the wave functions have negative exponentials
and their amplitude decreases with propagation.  These two roots are 
opposite to each other if $ c$ is negative.
 The propagation modes depend on confinement, magnetic field,
 Landau-level index, and electric field. For a given energy,
 $  g'( k_ { n\pm  }) =   (\hbar^{2} \Omega^{2} k_{n \pm}-
 \omega_{c}\hbar qE_{\perp})/m \omega^{2}_{T} $
and the replacement of the sum over $ k_{x} $ by 
an integral, $
               \sum_{k_{x}} \rightarrow  
	(L/2 \pi) \int^{L/2}_{-L/2} d k_{x}$,
 lead to
     \begin{eqnarray}
       \stackrel{\leftrightarrow}{\sigma}^{\epsilon_{p}}_{d} ({\bf r},{\bf r}') 
&=& 
                            \sum_{n}^{\epsilon_{p}} 
                            \frac{L}{2 \pi} \int^{\frac{L}{2}}_{-\frac{L}{2}} d k_{x}
                            \left [\frac{\delta ( k_{x} - k_{n+} ) }{| 
                            g'( k_{ +}) |}
                           +\frac{\delta ( k_{x} - k_{n-} ) }{| g'( 
                           k_{ - }) |} \right ]
                             \left ( \tilde{ \Lambda}^{-1} j({\bf r}) \right )_{ss} j({\bf r}')_{SS}
                             \nonumber \\
                             \nonumber \\
                        &=& \frac{L}{2 \pi} \sum_{n_{s}}^{\epsilon_{p}} 
                            \left [M_{k_{n+}} +   M_{k_{n-}}\right ] ,
                            \label{eq:dudu}
     \end{eqnarray}
 where
 \begin{equation}
   M_{k_{n\pm}} = \frac{1}{| g'( k_{ n \pm }) |}
   j({\bf r}')_{S\pm S\pm}
   \left (\tilde{ \Lambda}^{-1} j({\bf r}) \right )_{s\pm s\pm};
 \end{equation}
the notation $ s\pm$ or $S\pm$ indicates that only the values 
$k_{n\pm}$ are involved in the relevant $X_{ss'}$ or $X_{SS'}$ matrix 
element.

 \section{New conductance expressions in terms of transmission\\ and reflection
coefficients}

 \subsection{ Scattering formulation}
 For clarity the two roots $ k_{n\pm} $ are assumed to be in opposite directions.
 This holds if $ ac$ is negative and it is the case when the Hall field is neglected.
 Then Eqs. (\ref{cond}) and (23) give
 
          \begin{equation}
              G(\epsilon_{p}) =  \frac{L}{2 \pi}
                                 \sum_{n}^{\epsilon_{p}} 
                                 \left (N_{k_{ n+}}  +  N_{k_{ n-}} \right ),
                              \label{eq:roru}                                                                                     
          			\end{equation}
 where
         \begin{equation}
              N_{  k_{ n\pm}} = \frac{1}{| g'( k_{ n \pm }) |}
              \int dA'  j({\bf r}')_{ S\pm S\pm}
              \int dA 
              \left ( \tilde{ \Lambda}^{-1} j({\bf r}) 
              \right )_{s\pm s\pm}.
              \label{eq:seba}
         \end{equation}
 We now proceed with the evaluation of these two integrals that are 
 related to transmission and reflection coefficients.  We can carry 
 out the integrations  by choosing two 
 surfaces $A$ and $A'$ in an asymptotic region.  The 
 choice of surface is arbitrary.  It is not necessary to know the 
 exact scattering states.  It is sufficient to have their asymptotic 
 expression in a region away from the scattering centers.  The 
 scattering states are represented by a linear combination of 
 eigenfunctions of the unperturbed Hamiltonian.   
 The results for the various regions are  
 
 \begin{eqnarray} 
  {\overline{\psi}}_{n+} &=& \sum^{\epsilon_{p}}_{n'}
                                    t^{L}_{nn'}({\epsilon})
                                    \overline{\varphi}_{n'+}({\bf r}) ,
                                    \qquad  x \gg L_{s} , 
\label{eq:psi1a}\\
  {\overline{\psi}}_{n+} &=& \overline{\varphi}_{n+}({\bf r})
                                   +\sum^{\epsilon_{p}}_{n'}
                                   r^{L}_{nn'}(\epsilon_{p})
                                   \overline{\varphi}_{n'-}({\bf r}),
                                   \qquad x\ll 0 ,\label{eq:psi2a} \\
  {\overline{\psi}}_{n-} &=& \overline{\varphi}_{n-}({\bf r})
                                   +\sum^{\epsilon_{p}}_{n'}
                                   r^{R}_{nn'}(\epsilon_{p})
                                   \overline{\varphi}_{n'+}({\bf r}),
                                   \qquad x \gg L_{s},  
\label{eq:psi1b} \\
  {\overline{\psi}}_{n-} &=& \sum^{\epsilon_{p}}_{n'}
                                   t^{R}_{nn'}(\epsilon_{p})
                                   \overline{\varphi}_{n'-}({\bf r})
                                   \label{eq:pi2b}
                                   \qquad x \ll 0.
   \end{eqnarray}
 Using the normalization of the flux the current density is
 
\begin{equation}
          {\bf j}_{ \beta\alpha} = 
	 \sqrt{v_{\beta}v_{\alpha}}  \lambda 
                                   \overline{\psi}^{ *}_{\beta}            
                                   \stackrel{\leftrightarrow}{D} 
                                   \overline{\psi}_{\alpha},
         \end{equation}
where $\lambda =-iq \hbar/2mL$. %
Specifically for the different regions we have
 
\begin{eqnarray} 
  j_{{\overline{\psi}}_{n+}}({\bf r}')&=&  \lambda  v_{n+} 
                          \sum^{\epsilon_{p}}_{n'}
                          \sum^{\epsilon_{p}}_{n"}t^{L*}_{nn'}t^{L}_{nn"}
                          \overline{\varphi}_{n'+}
                          \stackrel{\leftrightarrow}{D} 
                          \overline{\varphi}_{n"+}, \qquad  x \gg L_{s}, \\
  j_{{\overline{\psi}}_{n+}}({\bf r}')&=& \lambda v_{n+} 
                          \{
                          \overline{\varphi}^{*}_{n+} 
                          \stackrel{\leftrightarrow}{D} 
                          \overline{\varphi}_{n+}
                       +  \sum^{\epsilon_{p}}_{n'}r^{L*}_{nn'} 
                          \overline{\varphi}^{*}_{n'-} 
                          \stackrel{\leftrightarrow}{D}
                          \overline{\varphi}_{n"+} \nonumber \\	           
                      &+& \sum^{\epsilon_{p}}_{n''}r^{L}_{nn"} 
                          \overline{\varphi}^{*}_{n'+} 
                          \stackrel{\leftrightarrow}{D}  
                          \overline{\varphi}_{n"-}  + \sum^{\epsilon_{p}}_{n'}                          
                         \sum^{\epsilon_{p}}_{n''}r^{L*}_{nn'}r^{L}_{nn"}
                          \overline{\varphi}^{*}_{n'-} 
                          \stackrel{\leftrightarrow}{D} 
                          \overline{\varphi}_{n"-}                             
                          \}, \qquad  x\ll 0,   \\
   j_{{\overline{\psi}}_{n-}}({\bf r}')&=& \lambda v_{n-} 
                          \{	               
                          \overline{\varphi}^{*}_{n-} 
                          \stackrel{\leftrightarrow}{D} 
                          \overline{\varphi}_{n-}
                        + \sum^{\epsilon_{p}}_{n'}r^{R*}_{nn'} 
                          \overline{\varphi}^{*}_{n'+} 
                          \stackrel{\leftrightarrow}{D}
                          \overline{\varphi}_{n"-} \nonumber \\	            
                     &+&  \sum^{\epsilon_{p}}_{n''}r^{R}_{nn"} 
                          \overline{\varphi}^{*}_{n'-} 
                          \stackrel{\leftrightarrow}{D}
                          \overline{\varphi}_{n"+} + \sum^{\epsilon_{p}}_{n'}                        
                          \sum^{\epsilon_{p}}_{n''}r^{R*}_{nn'}r^{R}_{nn"}             	               
                          \overline{\varphi}^{*}_{n'+} 
                          \stackrel{\leftrightarrow}{D} 
                          \overline{\varphi}_{n"+}                               
                          \}, \qquad  x \gg L_{s},   \\
     j_{{\overline{\psi}}_{n-}}({\bf r}')&=& \lambda v_{n-} 
                          \sum^{\epsilon_{p}}_{n'}                       
                          \sum^{\epsilon_{p}}_{n''}t^{R*}_{nn'}t^{R}_{nn"}                                
                          \overline{\varphi}^{*}_{n'-} 
                          \overline{\varphi}_{n"-},
                          \qquad x \ll 0.
		\end{eqnarray}

{\it Evaluation of the first integral}.
 Using Eqs. (\ref{eq:pda}) and (\ref{eq:prnd}) we obtain

\begin{eqnarray} 
   \int j_{{\overline{\psi}}_{n+}}({\bf r}') dA' &=& \frac{q  v_{n+}}{L}                         
   \sum^{\epsilon_{p}}_{n'} |t^{L}_{nn'}|^2,  \qquad  x \gg L_{s}, \\
   \int  j_{{\overline{\psi}}_{n+}}({\bf r}') dA' &=& \frac{q v_{n+}}{ L}                             
   \{ 1- \sum^{\epsilon_{p}}_{n'} |r^{L}_{nn'}|^2   \} \,  
     \qquad  x\ll 0, \\
   \int j_{{\overline{\psi}}_{n-}} ({\bf r}') dA'&=&  - \frac{q  v_{n-}}{L}                             
   \{ 1-\sum^{\epsilon_{p}}_{n'} |r^{R}_{nn'}|^2                            
   \}, \  \qquad  x \gg L_{s} \\
   \int j_{{\overline{\psi}}_{n-}} ({\bf r}') dA' &=& - \frac{q v_{n-}}{L}                               
   \sum^{\epsilon_{p}}_{n'}|t^{R}_{nn'}|^2, \qquad x \ll 0.
\end{eqnarray}
With flux conservation ($ 1= |r|^{2} + |t|^{2} $) we obtain the same result 
far away from each scattering region 
  \begin{equation}
         \int j({\bf r}')_{n\pm  n\pm  } dA'
        =\pm {qv_{n\pm}\over L}
         \sum_{s'}|t^{L(R)}_{nn'}|^2
         \label{eq:2int}
  \end{equation}
 
{\it Evaluation of the second integral}.
 The second integral has the superoperator $ \tilde{ \Lambda} $.  
 For elastic scattering it can be shown \cite{11d} that $ \tilde{ \Lambda} $ 
 has an exact inverse with dimension of time (=energy/$\hbar$). 
 We therefore write 
 $\tilde{ \Lambda} j({\bf r}) )_{s  s} = (1/\tau_{s} )  j_{ss}$ which leads to
 $( \tilde{ \Lambda}^{-1} j({\bf r}) )_{s  s} = \tau_{s}j_{ss}$ and
\begin{equation}          
\int(\tilde{ \Lambda} j({\bf r}) )_{ss}dA = \frac{1}{\tau_{s} }\int j_{ss}dA,
       \label{eq:tion}       
\end{equation}   
where $\tau_{s}$ is a characteristic time qualified below. 
We deduce the value of $ \tau_{s} $ as follows. Using Eqs. (18) and 
(36) we have
 \begin{equation}
               \beta_{\pm} = \int j_{n\pm n\pm } dA
                           = \pm (q v_{n\pm }/L ).     
		\label{eq:fraction}         
 \end{equation}
For the integral on the left-hand side of Eq. (\ref{eq:tion}), 
we use the result  \cite{10}
   \begin{equation}
      \left ( \tilde{ \Lambda} j({\bf r}) \right )_{n\pm n\pm  }
    =
 \frac{2\pi}{\hbar}
      \sum_{n'} \delta 
      ( \epsilon_{p} - \epsilon_{n'} ) | T_{n\pm  n'\pm} |^{2} 
      \left ( j_{n\pm n\pm } - j_{n'n'} \right ) ,
      \label{eq:ffff}
   \end{equation}
where $T_{n\pm n'\pm}=<\varphi_{n\pm }|V|\psi_{n'\pm  }>$ is the 
transition operator and $V$  the scattering 
potential. With $( \tilde{ \Lambda}^{-1} j({\bf r}) )_{s  s} = \tau_{s}j_{ss}$
 inspection of Eq.  (\ref{eq:ffff})
shows that $\tau_{s}$ is a characteristic time associated 
with the tunnel-scattering process. In the following though we will refer
to it simply as the characteristic time.

The Dirac $\delta$ function is rewritten in terms of the longitudinal components 
of the wavevector and of the two roots $k_{\pm }$. Then
 replacing   the
sum  over $k^{'}_{x}$ by an integral
leads to
 
\begin{equation}
       \left (  \tilde{ \Lambda} j({\bf r}) \right )_{n\pm   n\pm  } 
     = 
       \frac{L}{\hbar}
       \sum_{n'}^{\epsilon_{p}}  
       \left [ \frac{| T_{n\pm  n'+ }|^{2}}{|g'(k^{'}_{n'+})|}
                ( j_{nn} - j_{n'+} )
              +
              \frac{| T_{n\pm n'-}|^{2}}{|g'(k^{'}_{n'-})|}
              ( j_{nn} - j_{n'-}  ) \right ].
       \label{eq:hwhw}
   \end{equation}
\ \\
Using Eqs. (47), (48), and (50) the characteristic time becomes

\begin{equation}
\frac{1}{\tau_{n\pm }}= \frac{L}{\hbar}
                    \sum^{\epsilon_{p}}_{n'}
                    \left [ \frac{| T_{n\pm n'+} |^{2}}{|g'(k^{'}_{n'+})|}
                                   \left (1 \mp 
                                   \frac{\beta^{'}_{+}}{\beta_{n\pm }}   \right ) 
                              +
                  \frac{| T_{n\pm  n'-} |^{2}}{|g(k^{'}_{n'-})|}
                   \left (1 \pm \frac{\beta^{'}_{-}}{\beta_{n\pm }}\right ) \right ]. 
                               \label{eq:luc}
                  \end{equation}
\ \\
Using Eqs. (\ref{eq:2int}),  (\ref{eq:fraction}), 
(\ref{eq:luc}), and (\ref{eq:seba}) we get
%
 \begin{equation}
    N_{k_{n\pm}} = \frac{q^2}{L^2}
    \ \frac{ v_{n\pm}^{2}\tau_{n\pm}}
    {| g'( k_{n\pm }) |} 
                        \sum_{n'}|t_{nn'}|^2.
                        \label{eq:nuin}                   
 \end{equation}
  
\subsection{Evaluation of the conductance}
  
{\it Expression of $T_{ss'}$}. With $ V= h - h_{0} $ the matrix element 
$T_{ss'} = <\varphi_{s} | V | \psi_{s'} >  $
of  the transition operator $T$, between a 
state $\varphi_{s}$ and a scattering state $\psi_{s'}$,  becomes 

  \begin{equation}
    T_{ss'} = \epsilon_{s'} < \psi_{s} |\varphi_{s'} > - < 
\varphi_{s} | \left ( H_{0} | \psi_{s'} > \right ).
  \end{equation}
 We modify the second term on the right-hand side  so that the Hamiltonian 
operates on the left element. In order to do so we recall the expression
  
  \begin{equation}
      \int \varphi^{*} P_{x} \psi dv =
                                         \int P_{x} (\varphi^{*}  \psi) dv 
                                           + \int ( P^{*}_{x}  \varphi^{*}) \psi dv .
  \end{equation}
 With that we obtain 
 \begin{equation}
   \int \varphi^{*} P_{x} (P_{x}\psi)dv =
                                             \int (P^{2} \varphi^{*} )  \psi dv
                                               - \int  \frac{\hbar^{2}\partial}{\partial x} 
                                                 \left [ \varphi^{*} \frac{\partial}{\partial x}  \psi
                                                       - 
                                             \psi  \frac{\partial}{\partial x} \varphi^{*}  \right ] dv .    
  \end{equation}
If we combine these results with the Hamiltonian given
 by Eq. (\ref{eq:hamiltonien}) we obtain
 
 \begin{equation}
    < \varphi_{s} | ( H_{0} | \psi_{s'} > ) =
    (< \varphi_{s} | H_{0}|)
         \psi_{s'} >  -\frac{\hbar^{2}}{2m}
        \int\nabla (\varphi^{*} 
         \stackrel{\leftrightarrow}{\nabla} 
         \psi)dv  -\frac{qB}{m}
        \int P_{x}  (\varphi^{*} y \psi)dv . 
\end{equation}
 If we combine this result with Green's theorem, we obtain
   \begin{equation}
       T_{ss'} = (\epsilon_{s}-\epsilon_{s'})<\varphi_{s} | 
	\psi_{s'}>  
       + \frac{\hbar^{2}}{2m} \int_{A} d{\bf A}(\varphi^{*}_{s} 
         \stackrel{\leftrightarrow}{\nabla} \psi_{s'})
       + \frac{qB}{m}\int P_{x} (\varphi^{*}_{s} y  \psi_{s'})dv
    \end{equation}
 The first term is zero if the energies are the same. If so, the
 remaining terms can be simplified. The result can be written  compactly as

 \begin{equation}
       		T_{ss'} = \frac{\hbar^{2}}{2m} 
                 \int_{A} d{\bf A}\cdot \hat{{\bf x}}
                 \varphi^{*}_{s} 
                 (\stackrel{\leftrightarrow}{D}) \psi_{s'} .
 \end{equation}
 Finally, if we write it in terms of the normalized flux, we 
obtain
 \begin{equation}
       T_{ss'} = \frac{\sqrt{v_{s}v_{s'}}}{L} 
                 \frac{\hbar^{2}}{2m} 
                 \int_{A}  
                 d{\bf A}\cdot \hat{{\bf x}}
                 \bar{\varphi}^{*}_{s} 
                 (\stackrel{\leftrightarrow}{D})
                 \bar{\psi}_{s'}
    \end{equation}
  
 {\it $T_{ss'}$ in terms of transmission and reflection 
coefficients}.
  To evaluate the term $ T_{n\pm n'+} $, we use 
 Eqs. (\ref{eq:psi1a}) and (\ref{eq:psi1b}) together with 
Eqs.  (\ref{eq:pda}) and (\ref{eq:prnd}). 
%
For $x \gg L_s$ we obtain
\begin{equation}
  T_{n+n'+} = \frac{i\hbar}{L} \sqrt{v_{n+}v_{n'+}} t^{L}_{n'n}, 
\qquad  T_{n-n'+}= 0 .
\end{equation}
For $ x \ll 0$ the results are
\begin{equation}
    T_{n+n'+}=\frac{i\hbar}{L} \sqrt{v_{n+}v_{n'+}} \delta_{nn'},  
\qquad    T_{n-n'+}=-\frac{i\hbar}{L} \sqrt{v_{n-}v_{n'+}}r^{L}_{n'n} .
\end{equation}
%
To evaluate the term $ T_{n \pm n'-} $ we use Eqs. 
(\ref{eq:psi2a}) and   (\ref{eq:pi2b}) together with Eqs.
(\ref{eq:pda}) and (\ref{eq:prnd}).  %
For $x  \gg L_s$ we obtain
 \begin{equation}
     T_{n+n'-}= \frac{i\hbar}{L} \sqrt{v_{n+}v_{n'-}}r^{R}_{n'n},  
\qquad     T_{n-n'-}=-\frac{i\hbar}{L} \sqrt{v_{n-}v_{n'-}}\delta_{nn'} .
  \end{equation}
and for $ x \ll 0$ 
\begin{equation}
     T_{n-n'-}=-\frac{i\hbar}{L} \sqrt{v_{n-}v_{n'-}}t^{R}_{n'n}
\qquad     T_{n+n'-}= 0 . 
  \end{equation}

{\it Characteristic time in terms of transmission and 
reflection coefficients}. With the form of $T$ and the characteristic time given by 
Eq. (\ref{eq:luc}), the results for the various asymptotic regions are as follows.
For $x \gg L_s$  we have $1/\tau_{n-}= 0$ and $1/\tau_{n+}\neq 0$. 
 For $ x \ll 0$ the results are $1/\tau_{n+}=0$ and $1/\tau_{n-}\neq 0$.
 These nonzero results are given by
 %
   \begin{eqnarray}
     \frac{1}{\tau_{n\pm }} &=& \frac{\hbar}{L}
                         \sum_{n'}^{\epsilon_{p}}  
                         \left  [ v_{n\pm}v_{n'+}
                                |t^{L(R)}_{nn'} |^{2}(1-b_{\pm})/|g'(k^{'}_{n'+})|\right.
                                \nonumber\\  
                          &+&
                          \left. v_{n\pm}v_{n'-} 
                             |r^{R(L) }_{nn'}  |^{2}
                             (1+b_{\pm})/|g'(k^{'}_{n'-})|\right]; 
                          \label{eq:trau}
   \end{eqnarray}  
here  $b_\pm=\beta_{n'\pm}/\beta_{n}$ and $+$ (-)  corresponds to 
$t^{L}, r^{R}$ ($t^{R}, r^{L}$).
 This is simplified by noticing that $ g'( {\bf k}) = 
 \vec{\nabla}_{{\bf k}} \epsilon ({\bf k}) = \hbar {\bf v} $  gives 
 $| g'( k_{n\pm})| = \hbar v_{n\pm}$. Then Eq. (\ref{eq:trau}) takes the simpler
 form 
   
   \begin{equation}
    \frac{1}{\tau_{n\pm }} = \frac{1}{L}
                         \sum_{n'}^{\epsilon_{p}}  
                          \left [ v_{n\pm}
                                |t^{L(R)}_{nn'} |^{2}(1-b_{\pm}) 
                          +
                           v_{n\pm} 
                                |r^{R(L) }_{nn'}  |^{2}(1+b_{\pm})\right ]. 
                          \label{eq:trau1} 
                          \end{equation}  
 We emphasize the importance of this result. To our knowledge, with the exception of Ref. \cite{10}
 for $B=0$, the transmission and reflection coefficients have not been
associated with actual scattering  time in the literature. Here, through a
Master equation approach  we have an {\it explicit} result, for finite $B$,
relating these coefficients to the characteristic time.
 
 {\it Expression for the conductance }
  Using  Eqs. (\ref{eq:nuin}), (\ref{eq:trau1}) and (\ref{eq:fraction}) we obtain
                \begin{equation}
                   N_{  k_{n\pm}} = \frac{ q^2v_{n\pm }}{L 
                  | g'(  k_{ n\pm}) |} %
                                      \ \frac{ \sum_{n'} |t_{nn'}|^2  
                                      }{\sum_{n'}  X(n,n')},
                                               \end{equation}
    where                                        
                       \begin{equation}    
                          X(n,n') =  |t^{L(R) }_{nn'} |^{2}
                         ( 1 -v_{n'+}/v_{n\pm })  
                         +  |r^{R(L) }_{nn'} |^{2}  
                               ( 1  + v_{n'-}/v_{n\pm}  ).  
                   \label{eq:xaxa}
                   \end{equation}
 With current conservation 
 $\sum_{ n'} \left ( |t^{L (R)}_{nn'} |^{2} + |r^{R(L)}_{nn'}|^{2}\right ) =1 $, 
 this becomes %
 
 \begin{equation}
    		 N_{  k_{n\pm}} = \frac{q^{2}}{L}  
                         \sum_{ n'}  |t^{L(R)}_{nn'}|^2/ 
                         [1+\sum^{\epsilon_{p}}_{n'}
                         Y^{RL}_{\pm}(n,n')],  
                        \end{equation} 
                     where
                     \begin{equation}   
                      Y^{RL}_{\pm}(n,n') =  ( |r^{R}_{nn'} |^{2} v_{n'-}  
                                -
                                |t^{L}_{nn'} |^{2} v_{n'+} )/v_{n\pm }.
 \label{eq:final}
 \end{equation}
 %
 %
 Equations (\ref{eq:roru}) and (32) give the conductance as

 \begin{equation}
  G(\epsilon_{p})=\frac{q^{2}}{h}\sum_{ n}^{\epsilon_{p}}
                      \left [ \frac{\sum_{ n'} |t^{L}_{nn'}|^2}  
                      {\sum_{ n'}Y^{RL}_{+}(n,n')} +\frac{\sum_{ n'} 
                      |t^{R}_{nn'}|^2}
                      {\sum_{ n'}Y^{LR}_{-}(n,n')}\right ] 
 \end{equation}

 This new conductance expression is more general than the two-terminal
expressions of the literature. This can be  easily appreciated by realizing
that it has the following  interesting features.

 i) It is simplified
considerably if we neglect the Hall field; then  $v_{n+}=v_{n-}$ and  the two
terms in the square brackets  become identical. The same holds in the absence of
the magnetic field.  Actually, for $B=0$   Eq. (66) takes the form of Eq.
(4.184) of Ref. \cite{10}. The only difference is that in Eq. (66)  the
transverse channels and confining potential are
{\it explicitly} specified  whereas in Ref. \cite{10} they are not.

 ii) For uncoupled channels, i.e., for
 $r_{nn'}=r_{nn'}\delta_{nn'}$ and $t_{nn'}=t_{nn'}\delta_{nn'}$,  
 Eq. (64) gives the multichannel  version of Landauer's result,
  for identical  terminals,
 
  \begin{equation}
  G(\epsilon_{p})=\frac{q^{2}}{h}\sum_{n }^{\epsilon_{p}}
                      \frac{ |t_{nn}|^{2}}{ |r_{nn} |^{2}}=
                      \frac{q^{2}}{h}\sum_{n }^{\epsilon_{p}}\frac{ 
                      T_{n}}{R_{n}}
                                \label{eq:sc} 
\end{equation}
To our knowledge this is the first expression that shows this expected \cite{3},
 but absent from the literature, limit in the presence of a magnetic field.

 iii)   It is interesting to contrast the
$B=0$ limit of Eq. (66) with the $B=0$ result of Ref. \cite{3}. In this case
$v_{n+}=v_{n-}$. Proceeding then as  in Ref. \cite{10}  we  may  replace
$1/v_{n'+}\propto \tau_{n}$  by $(1/N)\sum_{n'} (1/v_{n'})$  and
make an average over the channels to obtain the result of Ref. \cite{3},
i.e.,

 \begin{equation}
  G(\epsilon_{p})=\frac{q^{2}}{h} 
                     \sum_{ n}  T_{n} \frac{  \sum_{n}(2/v_{n})}
                      {\sum_{ n}  ( 1+ R_{n} -T_{n})/v_{n}}, 
 \end{equation}
 if we remember that $R_{n}=\sum_{n'} |r_{nn'} |^{2}$ and $T_{n}= 
 \sum_{n'} |t_{nn'} |^{2}$. Despite its approximate character the 
 procedure indicates that Eq. (66) is more general than  Eq. (68)
 even for $B=0$.

iv) For  $R\approx 1$ and $T\ll 1$,   Eq. (66) gives the standard
\cite{3}, \cite{7}
result $G(\epsilon)=(q^{2}/h) Tr\{tt^{*}\}$ if we assume a {\it weak} 
\cite{3}
channel coupling such that  $v_{n'}\ll v_{n},\ n'<n$.  

v) When the strength of the scattering is vanishingly
small, we have $r\approx 0$ and $t\approx 1$. As expected, in this case for 
identical terminals and $v_{n'}\ll v_{n}, n'<n$, the conductance diverges,
as realized in a four-terminal (two leads, two probes) experiment.

iv) Finally, we notice that the expression contains the Hall field, through the
factors $v_{n\pm}$, cf. Eq. (21), which accounts for the electron-electron
interaction in the Hartree sense \cite{11}.

\subsection{Conductance in tilted magnetic fields}

Equation (66) is valid for a perpendicular magnetic field
$B$ parallel to the z axis.  It is of interest to have an expression 
valid for tilted fields $B$ but the solution of Schroedinger's 
equation becomes very unwieldy and, to our knowledge, can be obtained 
only numerically when $B$ points in an arbitrary direction.  However, 
in one   particular case   a simple analytic solution exists and leads to a
generalization of the conductance (66). Below we briefly derive the relevant
expression since we are not aware of any pertinent result in the literature. 
This is the case when the field $B$ is in the (x,z) plane and has components
$B_{\parallel}$ along $\hat x$ and $B_{\perp}$ along $\hat{z} $. The situation
is described by the vector potential $ {\bf A} = B_{\perp}y\hat{x} +
B_{\parallel}y\hat{z}$. Assuming an eigenfunction $ \psi (x,y) = f(y) e^{i
k_{x}x} $ the Hamiltonian gives

\begin{equation}
   \left[\frac{\hbar^{2}k^{2}_{x}}{2m}-y(\omega_{\perp}\hbar 
k_{x}+qE_{\perp}) 
     + \frac{1}{2}m(\omega^{2}_{B} + \Omega^{2})y^{2}\right]f(y) 
   - \frac{ \hbar^{2}}{2m} f^{''}(y)  = \epsilon f(y) 
\end{equation}
With  $\xi = ( \omega_{\perp}\hbar k_{x} +qE_{\perp})/m\omega^{2}_{B}$ 
 this   equation is transformed to

\begin{equation}
  \frac{m\tilde{\omega}^{2}_{T}}{2}(y - \frac{\omega^{2}_{B}}{\tilde{\omega}^{2}_{T}} 
\xi)^{2}f(y)
    - \frac{\hbar^{2}}{2m} f^{''}(y) = E f(y), 
\end{equation}
 where $\tilde{\omega}^{2}_{T}=\omega^{2}_{B}+\Omega^{2} $ and 
 $\omega^{2}_{B}=\omega^{2}_{\parallel}+\omega^{2}_{\perp}$. 
 This is again an equation for a (displaced) harmonic oscillator. 
 The corresponding eigenvalues $\epsilon \equiv \epsilon( k_{x}, n)$ are  

\begin{equation}
 \epsilon(k_{x}, n)=(n+1/2)\hbar\tilde{\omega}_T - 
  \left[\hbar^2k_x^2(\Omega^{2} +\omega^{2}_{\parallel})
 -2qE_{\perp}\omega_{\perp}\hbar k_x -q^2 E_{\perp}^{2}\right]\big /2m\omega^{2}_{T}
\end{equation}

As can be seen these results are  similar to those obtained 
when the field  $B$ is parallel to the z axis. In fact,  Eq. (13)
can be obtained from  Eq. (71) by  setting
 $B_{\parallel}=0$ which entails  $\omega^{2}_{B}=\omega^{2}_{c}$.  
All the analysis of Sec. III can be repeated and the result
for the conductance has the same form. The only thing that changes
in Eq. (66) are the roots $k_{n\pm}$, cf. Eq. (26); they now involve
Eq. (71) rather than Eq. (13).
As a side remark we notice that in a longitudinal magnetic field, with 
$ B_{\perp}=0 $, we obtain formally the same result  as in the absence of the
magnetic field since the carriers are free in a parallel magnetic field. 

\section{Discussion}

 The expression for the conductance, given by Eq.  (66), is very
 general and not limited to two identical terminals.  We can 
 interchange the indices $R$ and $T$ without changing the expression.  
 This means that the conductance does not depend on the direction of 
 the current. This and the various limits   this expression reproduces  show 
 its generality.

 This result for the conductance, valid when a magnetic field is 
 present, was not anticipated in Ref.  \cite{10}.  Since at first 
 sight in a magnetic field the eigenfunctions along two opposite 
 directions would be separated by a distance $\propto 2k_{x}$, it was thought 
 that the expression 
 would change dramatically.  As shown though, incorporating directly 
 in the one-electron Hamiltonian the magnetic field, the (parabolic) 
 confining potential, and the Hall field, lead to an eigenfunction suitable 
 for the calculations.  It showed explicitly the cancellation 
 in the product of the velocity with the quasi-one-dimensional density
 of states in the current carried by a channel or mode and simplified 
 the final result.  In addition, it allowed the consideration of 
 tilted magnetic fields (in the (x,z) plane) and of the 
 electron-electron interaction in a mean-field or Hartree sense since 
 the Hall field was taken constant across the width whereas it is not 
 since its value near the edges is different than that in the main part of 
 the sample \cite{11}. The  last  two aspects,   limit ii) of Eq. (66),
 and Eq. (61) for the characteristic time are missing  from other
expressions for  the magnetoconductance  \cite{3,4,7,12}.
The most common general formula   \cite{7} reads  $
 G_{mn}(\epsilon) = (q^{2}/h) \sum^{\epsilon}_{ac} 
 |t_{mn,ac}|^{2}$, where $t_{mn,ac}$ is the transmission coefficient between channel $a$ 
in terminal $m$ and channel $c$ in terminal $n$.  
This  formula  applies to a multiterminal configuration and  two-probe 
 measurements \cite{5} whereas ours applies to a two-terminal configuration.

 \acknowledgements
 
 We thank Dr. F. Benamira for  fruitful discussions. 
This work  was supported by the Canadian NSERC Grant No. OGP0121756.

\begin{figure}
\Large
\begin{picture}(400,200)
    \put(40,55){\line(0,1){50}}
    \put(10,90){$dS_{L}$}
    \put(35,80){\vector(-1,0){20}}
    \put(70,105){\vector(0,-1){50}}
    \put(47,75){$\vec{E_{\perp}}$}
    \put(80,90){i}
    \put(85,90){\vector(1,0){75}}
    \put(80,75){j}
    \put(160,75){\vector(-1,0){75}}
    \put(100,62){$R_{ij}$}
    \put(160,55){\framebox(80,50)}
    \put(170,120){\vector(1,0){60}}
    \put(195,125){$\vec{E_{x}}$}
    \put(75,120){\vector(1,0){10}}
    \put(90,117){x}
    \put(75,120){\vector(0,1){10}}
    \put(73,140){y}
    \put(90,140){$\otimes$}
    \put(105,140){$\vec{B_{z}}$}
    \put(220,45){\vector(1,0){20}}
    \put(195,40){$L_{s}$}
    \put(180,45){\vector(-1,0){20}}
    \put(45,105){\line(1,0){310}}
    \put(45,55){\line(1,0){310}}
    \put(195,80){S}
    \put(240,80){\vector(1,0){75}}
    \put(320,80){j}
    \put(265,90){$T_{ij}$}
    \put(330,55){\line(0,1){50}}
    \put(335,90){$dS_{R}$}
    \put(335,80){\vector(1,0){20}}
     \put(210,90){\circle*{5}}
     \put(165,100){\circle*{5}}
     \put(200,60){\circle*{5}}
     \put(187,85){\circle*{5}}
     \put(233,77){\circle*{5}}
     \put(170,65){\circle*{5}}
     \put(230,95){\circle*{5}}
     \put(186,71){\circle*{5}}
     \put(200,97){\circle*{5}}
     \put(207,70){\circle*{5}}
     \put(175,85){\circle*{5}}
     \put(220,65){\circle*{5}}
     \put(220,79){\circle*{5}}
%
%
%
%
    \put(195,18){L}
    \put(158,25){\vector(-1,0){120}}
    \put(238,25){\vector(1,0){120}}
    
\end{picture}
\caption{A quasi-one-dimensional conductor, connected to left (L) and 
right (R) reservoirs in the presence of crossed electric and a 
magnetic fields.  The length of the conductor is L. The solid dots 
represent random scattering centers.}
\end{figure}
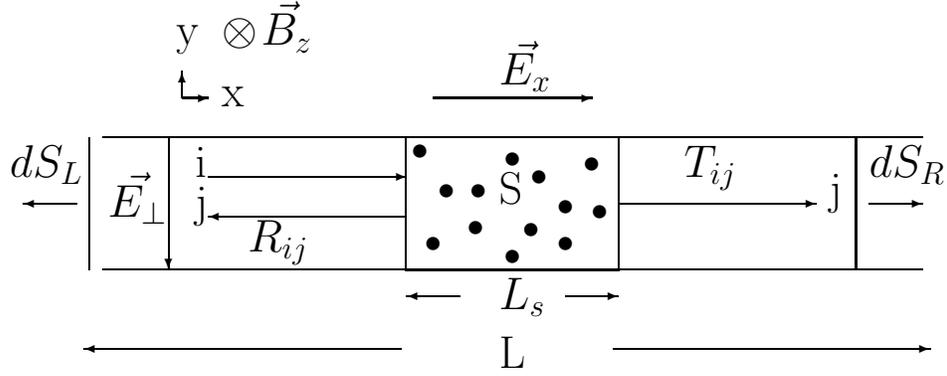
 

\begin{thebibliography}{abc}
\bibitem[\dagger]{sguill}  E-mail: guillon@canr.hydro.qc.ca
\bibitem[\diamond]{takis}  E-mail: takis@boltzmann.concordia.ca
 
\bibitem[\diamondsuit]{vanvliet} E-mail: vanvliet@physics.miami.edu
 
\bibitem{1} B. J. Van Wees, H. Van Houten, C.W.J. Beenakker, J. G. 
Williamson, L. P. Kouwenhoven, D. Van der Marel, and C. T. Foxon, Phys. Rev. 
Lett. {\bf 60}, 848 (1988); D. A. Wharam, T. J. Thornton, R. Newbury, M. Pepper, H. 
Ahmed, J. E. F. Frost, D. G. Hasko, D. C. Peacock, D. A. Ritchie, and 
G. A. C. Jones, J. Phys. C: Solid State Phys. {\bf 21}, L209 (1988)
\bibitem{2} R. Landauer, IBM J. Res. Dev. {\bf 1}, 223 (1957); Phil 
Mag.  {\bf 21}, 863 (1970). 

\bibitem{3} M. Ya. Azbel, J. Phys. C: Solid State Phys. {\bf 14}, 
L225 (1981); M. Buttiker, Y. Imry, R. Landauer and S. Pinhas, Phys. 
Rev. B {\bf 31}, 6207 (1985).

\bibitem{4} D. C. Langreth, E. Abrahams, Phys. Rev. B {\bf 24}, 
2978 (1981); J. Kucera and P. Streda, J. Phys. C: Solid State {\bf 
21}, 4357 (1987);  K. Shepard, Phys. Rev. B {\bf 43}, 11623 (1991);
J. U. Nockel and al, Phys. Rev. B {\bf 48}, 17569 (1993).

\bibitem{4a}E. N. Economou and C. M. Soukoulis, Phys. Rev. Lett. {\bf 46}, 
 618 (1981); D. S. Fisher and P. A. Lee, Phys. Rev. B {\bf 23}, 6851 (1981).

\bibitem{4b} D. J. Thouless, Phys. Rev. Lett. {\bf 47} 972 (1981);
E. N. Engquist and P. W. Anderson, Phys. Rev. B {\bf 24}, 1151 (1981);
A. D. Stone and A. Szafer, IBM J. Res. Dev. {\bf 32}, 
384 (1988).

\bibitem{5} Y. Imry, {\it Introduction to Mesoscopic Physics}, Oxford 
University Press, (1997)

\bibitem{6} C. W. J. Beenakker and H. Van Houten, Solid State Phys. 
{\bf 44}, 1 (1991).

\bibitem{7} H. U. Baranger and A. D. Stone, Phys. Rev. B {\bf 40}, 8169 (1989).

\bibitem{8} M. Buttiker, Phys. Rev. Lett. {\bf 57}, 1761 (1986).

\bibitem{9} A. D. Stone, Phys. Rev. Lett. {\bf 54}, 2692 (1985);

\bibitem{9a} D. A. Benoit {\it et al.}, Phys. Rev. Lett. {\bf 57}, 1765 (1986).

\bibitem{10} F. Benamira, Ph. D. Thesis, University of Montreal (1996).

\bibitem{11} A. H. Macdonald, T. M. Rice, and W. F. Brinkman, Phys. 
Rev. B {\bf 28}, 3648 (1983). 

\bibitem{11a} C. M. Van Vliet, {\it Quantum Transport in Solids},
CRM Proceedings and Lecture Notes, vol. 11, 421-449, The
Am. math. Society, 1997. 

\bibitem{11c} O. Heinonen, P. L. Taylor, and S. M. Girvin, 
Phys. Rev. B  {\bf 30}, 3016 (1984); O. G. Balev and P. Vasilopoulos,  
{\em ibid}  {\bf 47}, 16410 (1993).

\bibitem{11d} C. M. Van Vliet, {\it Statistical Thermodynamics II,
Nonequilibrium Statistical 
Mechanics}, Lecture Notes, Part E: Universite de Montreal, 
CRM-1618 (1989), unpublished.

\bibitem{12}  P. Streda, J. Kucera, and A. H. Macdonald, Phys. Rev. Lett. {\bf 17}, 
1973 (1987); J. K. Jain, Phys. Rev. B {\bf 37}, 4276 (1988).


\end{thebibliography}
\end{document}